# Development of a Machine Learning Potential to Study Structure and Thermodynamics of Nickel Nanoclusters

Suvo Banik[1,2], Partha Sarathi Dutta[1,2], Sukriti Manna[1,2], Subramanian KRS Sankaranarayanan*[1,2]

[1] Center for Nanoscale Materials, Argonne National Laboratory, Lemont, Illinois 60439.

[2] Department of Mechanical and Industrial Engineering, University of Illinois, Chicago, Illinois 60607.

*skrssank@uic.edu

**Abstract:**

Machine Learning (ML) potentials such as Gaussian Approximation Potential (GAP) have demonstrated impressive capabilities in mapping structure to properties across diverse systems. Here, we introduce a GAP model for low-dimensional Ni nanoclusters and demonstrate its flexibility and effectiveness in capturing the energetics, structural diversity and thermodynamic properties of Ni nanoclusters across a broad size range. Through a systematic approach encompassing model development, validation, and application, we evaluate the model's efficacy in representing energetics and configurational features in low-dimensional regimes, while also examining its extrapolative nature to vastly different spatiotemporal regimes. Our analysis and discussion shed light on the data quality required to effectively train such models. Trajectories from large scale MD simulations using the GAP model analyzed with data-driven models like Graph Neural Networks (GNN) reveal intriguing insights into the size-dependent phase behavior and thermo-mechanical stability characteristics of porous Ni nanoparticles. Overall, our work underscores the potential of ML models which coupled with data-driven approaches serve as versatile tools for studying low-dimensional systems and complex material dynamics.

**Introduction:**

Metallic nanoparticles (MNPs) possess fascinating chemo-physical properties due to their high surface-to-volume ratios, which are vastly different from those of individual atomic constituents and their bulk counterparts[1-3]. These zero-dimensional systems, owing to the variability in atomic packing and the presence of isomorphism, entail a diverse energy landscape through non-trivial interplay between their structure and energetics [4-6]. The properties of MNPs strongly depend on their structural and morphological features, such as their size, shape, chemical composition, and chemical ordering[7,8]. Therefore, harnessing the full potential of MNPs hinges on our understanding of the energy landscape and its mapping to the corresponding structure.

Nickel (Ni) is abundant in the Earth's crust, making it more cost-effective than most of the metals used in various applications. Nickel nanoparticles (Ni NPs), owing to their unique morphology, crystallinity, and specific surface area (SSA), possess magnetic, chemical, and physical properties with potential applications in areas such as high-temperature filters, ink for nanotube printing, lithium-ion batteries, textiles, optical switches, and in immobilization of biomolecules[9-14]. Compared to other magnetic nanoparticles, Ni NPs possess superior catalytic properties which makes them useful in reactions, as propellants, sintering additives in coatings, plastics, fibers, etc[12,14]. One way to further improve the performance and applicability of nanoparticles is through the introduction of porosity. Porous nanoparticles yield a unique combination of physical and morphological traits enabling greater control of the local chemical environment[15]. For example, Schwartzberg et al.[16] reported the synthesis, characterization, and tunable optical properties of hollow gold nanoparticles, which have shown great potential for chemical and biological sensing applications. Additionally, hollow nanoparticles (hNP) exhibit remarkable performance in areas such as catalysis, plasmonics, sensing, and energy storage, among others[15,17,18]. Thus, understanding the synthesis routes of these materials becomes necessary, which in turn necessitates an understanding of their thermal stability, energetics, and dynamics. Knowledge of the shape and size dependent nature of stability would not only entail finer control of their performance but can also make synthesis cost-effective. For example, maximizing the hNP size while decreasing the shell thickness is an alternative to produce lower-cost and lightweight devices and improve their performance for storage, diffusion, or other surface-activated processes[17].

Numerical simulations, such as Molecular Dynamics (MD) and ab initio methods, can unveil phase stability and transition mechanisms of nanoparticles at an atomistic level[13,19-21], offering insights into synthesis pathways. While simulations relying on Density Functional Theory (DFT) are accurate[22], they are computationally expensive which limits their access to often large spatiotemporal scales crucial to understand the structure and dynamical evolution of these systems. Semi-empirical potentials extend accessibility to length and timescales but lack flexibility. Thus, the eventual choice of model entails a trade-off between computational speed and accuracy. A challenge is the fixed functional forms with few adjustable parameters in these potentials, when fitted to bulk systems, can lead to significant errors in predicting low-dimensional systems with relatively smaller sizes[23]. This issue is exacerbated by the strong correlation between energetics and stability in low-dimensional systems and with system size, often overlooked when extrapolating from bulk data[23].

Machine Learning Force Fields (ML-FFs)[24-30] offer a promising alternative, with computational speed between empirical models and ab initio methods, achieving accuracy on par with DFT in mapping the energy landscape of nanoclusters. ML potentials surpass the size limitations of DFT and can capture nanoparticles' energy landscape directly from data, without relying on predetermined functional forms. However, this necessitates training data to have unique and representative minima of the energy landscape spanning across a wide range of energetics. While ML potentials can efficiently emulate ab initio methods in yielding longer trajectories containing information on phase change mechanisms and transition pathways, characterizing them deterministically also presents a challenge. The implementation of mathematical formulation-based descriptors, such as order parameters and the Lindemann index[19], is mostly based on empiricism and/or intuition. Moreover, insights into local atomic environments in nanoclusters from these descriptors are limited and become fuzzier as system size decreases[31,32]. Data-driven approaches such as Graph Neural Networks (GNN) have proven to provide insightful characterization based on features learned solely from data[33,34], eliminating the need for chemical intuition and heuristic approaches.

Thus, for modeling low dimensional systems such as nanoclusters, the development of ML potential models and subsequent characterization are indispensable tasks. Regarding the former, it's noteworthy that while numerous studies have focused on developing ML potentials for bulk systems[26-30], only a handful of them have been conducted for nanoclusters[25,35,36]. Additionally, the datasets used in these studies are typically limited to a few size-dependent clusters, primarily focusing on relatively large nanoclusters (- nanoparticles) (>100 atoms)[35-37]. In contrast, there are considerable variations in morphology and energetics in the lower-size regions (<50 atoms)[38,39]. Therefore, from a perspective of generalized application, it is imperative to assess the ability of ML potentials to accurately capture the energy landscape of nanoclusters in these regimes through rigorous validation protocols. It is also essential to determine the extrapolative nature of these cluster potentials as the system size increases and the bulk nature of properties begins to manifest. Such investigations will not only aid in further developing these potentials but also provide insights into the size and diversity requirements of the data but also shed light on the strengths and limitations of these potentials across various operational regimes.

In this study, we develop an ML potential model namely the GAP (Gaussian Approximation Potential)[26-30,35,40] for Nickel using an extensive cluster training dataset spanning from 2 to 55 atoms. We comprehensively evaluate the predictability, flexibility, and accuracy of our model across different size regimes of nanoclusters. The validation process involves assessing the model's capability to capture DFT local minima across various sizes in terms of both energetics and structural attributes, comparing vibrational modes, and evaluating dynamic stability across a multitude of sizes and configurations. To assess the extrapolative nature of our potential, we calculate bulk properties (e.g., lattice parameters, melting) and examine our model's ability to capture size-dependent melting. We utilize a Graph Neural Network-based architecture, CEGANN[33], to gain insights into the evolution of the local environment during nanoparticle melting. By leveraging our parameterized GAP model and the CEGANN classification workflow, we also investigate the intriguing thermal stability and temperature-induced structural changes in Ni hollow nanoparticles.

**Result:**

**The GAP model:**

The overall workflow for fitting the Gaussian Approximation Potential (GAP) for Ni nanoclusters is illustrated in Fig. 1(a). The dataset consists of nanoclusters ranging in size from 2 to 55 atoms (comprising 2085 configurations), encompassing a broad energy spectrum (refer to the dataset section). The training data, which includes energies and forces of the clusters, was computed using Density Functional Theory (DFT) (refer to supplementary Note 1) generated *via* various sampling techniques such as Nested Ensemble and evolutionary approach. We utilize the GAP framework, which employs sparse Gaussian Process Regression (GPR) customized for fitting Potential Energy Surfaces (PESs) and local properties of atomic systems. A fundamental assumption in fitting the PES of atomic systems is that the total quantum mechanical energy can be decomposed into local contributions ε that depend on descriptors[40]

$$E = \sum_d^{\text{descriptors}} \sum_{i=1}^{N_d} \varepsilon_d$$

where $N_d$ is the number of descriptors of type d. Descriptors, such as the Smooth Overlap of Atomic Positions (SOAP)[41], can serve as arguments to n-body energy terms based on interatomic distance. The SOAP kernel $K(R_i, R_s)$[40,41] characterizes the similarity between two neighborhoods: it reaches its maximum when the two neighborhoods are identical. In the Gaussian Approximation Potential (GAP) framework developed by Csanyi and coworkers, each energy term $\varepsilon_d$ is expressed as an independent sparse Gaussian Process in the form[40]

$$\varepsilon_d(X) = \sum_{m=1}^{M_d} c_m k_d(x, x_m)$$

where $M_d$ is the number of sparse or representative points of descriptor d, $k_d$ represents the kernel, covariance, or similarity function, and $c_m$'s are the fitting coefficients obtained through the dataset configurations. The target properties, denoted by y, of the fitting procedure, are therefore the sum of local energy contributions in the form of total energy. We implement the MAML[42] package for our work to fit the GAP model (refer to the method section on the GAP model).

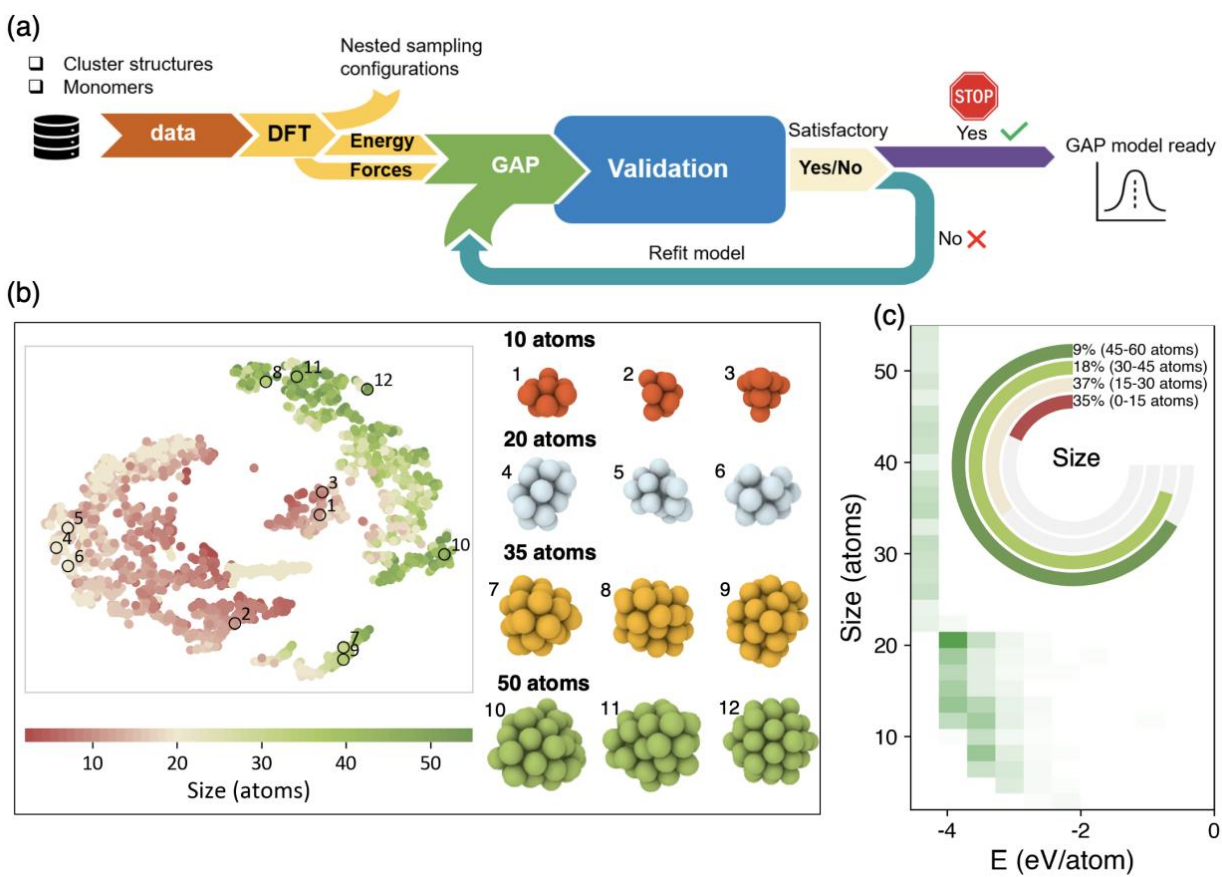

*Fig. 1. Development of a Gaussian Approximation Potential (GAP) model for Ni based on first-principles dataset of cluster energetics and forces. (a) Schematic workflow showing fitting and validation of the GAP model. (b) Shows the t-SNE plot of the SOAP representation of the cluster dataset used for fitting the GAP potential and corresponding size-dependent structural variability of the clusters. (c) Size-dependent energetics and size distribution of the Ni clusters in the dataset.*

**The cluster dataset:**

The accuracy of ML potentials relies heavily on training datasets that accurately depict the intricacies of the energy landscape, encompassing various local minima and maxima. We adopt three strategies to sample the configurations and generate the data: (a) Using Nested Ensemble-Based Search for Entropically Favored Clusters: A Boltzmann-based Metropolis sampling and a nested ensemble-based approach[43] were employed to generate the structures (refer to supplementary Note 2). (b) Structure Mining for Low-Energy Cluster Configurations. Additionally, data were mined from databases such as the Quantum Cluster Database (QCD)[44] and the Cambridge Cluster Database[45]. A crucial aspect of generating the cluster dataset for nickel clusters during DFT calculations is the consideration of multiple magnetic configurations. Clusters can often become trapped in different local minima due to variations in magnetic configurations. Numerous studies on nickel nanoclusters consistently show that both the clusters and their isomers ferromagnetic structures[44,46,47], including configurations found in databases like QCD[44]. We employed a systematic approach (detailed in Supplementary Note 1) to compute the energy and forces during DFT calculations while accounting for the magnetic configurations, which ultimately led to all configurations converging to a ferromagnetic state. A total of 2085 cluster configurations were considered in these calculations, which were later split into an 80:20 ratio for training and testing purposes. For model validation, we generated an additional independent, uncorrelated dataset using an EAM potential model[48] in combination with the CASTING framework[1]. This allowed us to sample a total of nickel clusters, ranging from 5 to 30 atoms, across a wide energy window. We also included Ni clusters from the literature1. In total, 290 samples were generated. We then evaluated the energy and forces of these configurations using the same protocol applied earlier.

.

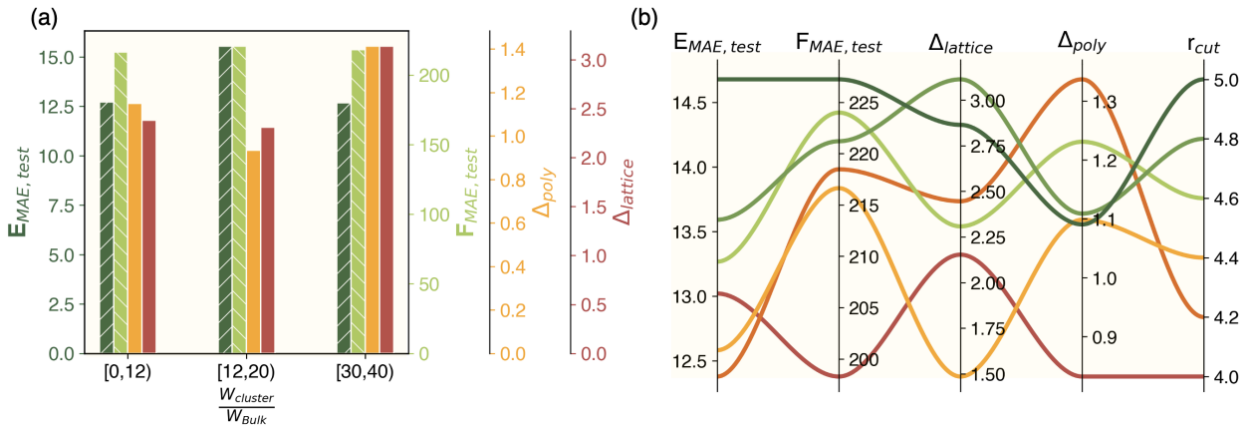

*Fig. 2. Selecting the optimal GAP model for describing atomistic interaction in Ni. (a) The mean performance of GAP models for different ranges of weights of cluster to bulk data used during training. The performance metrics include energy error ($E_{MAE,test}$ -meV/atom) and force error ($F_{MAE,test}$ -meV/Å) on clusters of the test dataset, as well as errors in cohesive energy and ordering ($\Delta_{poly}$) and lattice parameters ($\Delta_{lattice}$) for predicting low energy polymorphs of bulk Ni. (b) The mean performance metrics as in (a) for different values of cutoff ($r_{cut}$) used during training the models.*

A reduced dimensional representation of training and test cluster dataset sampled across various sizes is outlined in Fig. 1 (b-c). As shown in Fig. 1 (b), the t-SNE plot of SOAP representation of the clusters reveals diverse and distinct structural domains, particularly noticeable for small-sized clusters (<15 atoms), highlighting marked structural disparities compared to larger ones. With decreasing cluster size, a multitude of structural morphologies emerges due to heightened surface effects and quantum confinement[49,50]. Understanding the energy landscape poses a significant challenge due to its ruggedness, where even minor configuration adjustments lead to significant energy shifts and noticeable size-dependent energy variations, as evidenced in our dataset (Fig. 1 (c)). While the overall energy range falls within (-5 to 0 eV), the most pronounced energy fluctuations occur in smaller clusters (<20 atoms), occupying a broader space in the t-SNE plot, indicating heightened structural diversity and a rugged energy landscape. Approximately 35% of configurations belong to the small cluster size category, with an additional 37% falling into the medium-size category (15-30 atoms). However, the dataset faces a challenge due to its maximum configuration size of 55 atoms, where typical bulk characteristics may not manifest. To enhance the generalizability of the potential, we further included the Equation of State (EOS) data (63 configurations) of bulk FCC polymorphs in the dataset, capturing bulk attributes such as lattice parameters and melting behavior. Nonetheless, the dataset uniformly spans a wide energy window crucial for effectively capturing the Potential Energy Surface (PES) across a broad size range from clusters to bulk.

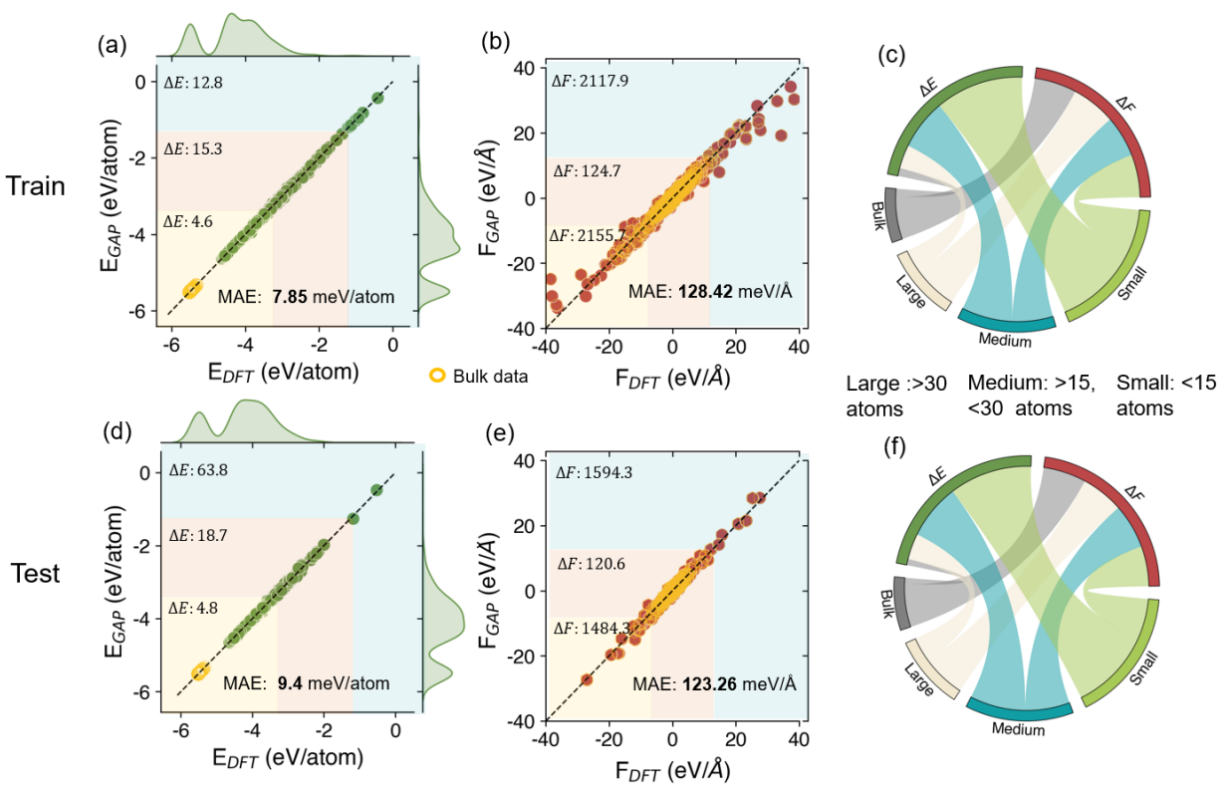

*Fig. 3. Performance of our parameterized GAP model for Ni. (a-b) Parity plot displaying the energy and force errors on the training dataset for the cluster data and included Bulk data. (c) Chord diagram displaying the contribution towards error (on energy E and force F) from 3 different size categories of the clusters (Large: >30 atoms, Medium: >15 and <30 atoms, and Small: <15 atoms) for the training dataset. (d-e) Parity plot displaying the energy and force errors on the test dataset. (f) Chord diagram displaying the contribution towards error from 3 different size categories for the test dataset.*

**Selection of the optimal model:**

A crucial part of the GAP workflow, as depicted in Fig. 1a and in the methods section, is the validation stage, which needs to assess the generalizability of the fitted GAP model in terms of its performance on both the clusters and the bulk attributes. Two crucial parameters to consider are the degree of weight given to the bulk EOS data and the cutoff for building the neighborhood during fitting the model itself (which determines locality of the model. i.e., short or a long-range). Fig. 2 depicts the mean performance of 114 GAP models influenced by these two crucial hyperparameters. It can be observed that for different emphasis on the bulk data, there is relatively lower variations in energy error ($E_{MAE,test}$) and force error ($F_{MAE,test}$ - meV/Å) on the clusters of the test dataset (Fig. 2a), around ~2 meV/atom, while there is a considerable increase in the errors predicting cohesive energy - ordering ($\Delta_{poly}$) and lattice parameters ($\Delta_{lattice}$) (refer to supplementary Note 3 for the calculation of these parameters) for predicting the bulk low energy polymorphs as the ratio of cluster to bulk data during training ($\frac{W_{cluster}}{W_{Bulk}}$) goes beyond 30 (Fig. 2a). The overemphasis on the clusters, although it keeps the energetics and the forces on the clusters intact, harms the bulk performance. At the same time the cutoff also influences these performance parameters significantly (Fig. 2b). For a relatively larger cutoff (5Å), there is improvement in $\Delta_{poly}$ at the sacrifice of energetics and the forces on the clusters. At the higher values (> 4.4 Å), the trends indicate a relatively high error on the performance metrics, while the overall performance is relatively better at lower range of cutoffs (Fig. 2b). This is primarily due to the over presence of small size clusters, making the overall ratio of surface atoms to volume atoms ratio higher in the dataset. In general for bulk systems having a large cutoff helps in achieving accurate predictions and capturing longer-range order, which is essential for numerous physical phenomena, such as thermodynamic phase behaviors[51]. However, this comes at the expense of increased computational cost for predictions from ML potential models.. For bulk systems, higher cutoffs may be feasible due to periodicity. Typically, the selected cutoff for ML potential models trained on bulk data is around ~4 Å[52]. Going beyond this for bulk systems may not significantly affect bulk performance; however, for clusters, regardless of the emphasis placed on the bulk EOS data, the upper range of the cutoff becomes crucial. A higher cutoff would incorporate more surface atoms in the local environment construction, leading to relatively higher errors. While it is intuitive that a too low cutoff will result in nearsighted potentials[51], namely, they lack knowledge of structure or interactions beyond the local atomic environment, causing deterioration of bulk properties such as melting. Thus, the choice of cutoff is crucial while building ML potential models from cluster datasets. For our case, we proceed with sets that satisfy a minimum value for all the above-mentioned criteria (refer supplementary Fig. 2), for further evaluating properties such as elastic constants, the Equations of State (EOS), and surface energies. The final model is selected afterwards based on the performance across these multiple objectives.

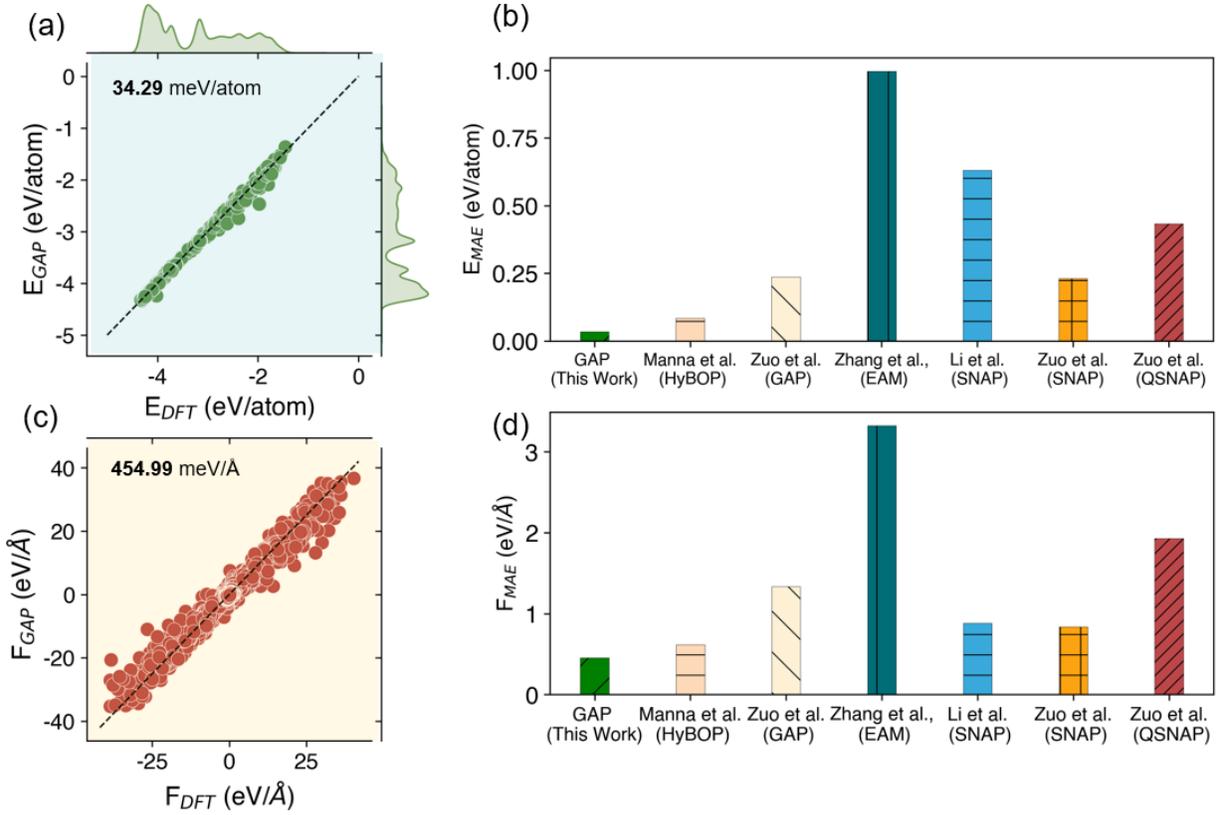

***Fig.4.*** *Validation of the GAP model on uncorrelated data sampled using the EAM model[48]: (a,c) Energy and force predictions of the GAP model compared to the DFT counterpart for different nanoclusters of Ni sampled. (b,d) Comparison of the performance of the GAP model with other potential models in terms of MAE error in predicting energy and forces as compared to DFT.*

**Model Performance:**

Initially, we determined, that a weight ratio of ~4:1 for cluster to bulk in the training set results in optimal performance in terms of obtaining both the cluster and bulk attributes in good agreement with DFT data, (refer to Supplementary Fig. 2 and Supplementary Note 3). The errors on the training and testing sets for the selected GAP model is depicted in Fig. 3. From the results on the training dataset (Fig. 3a), the overall energy error on the training dataset is 7.85 meV/atom. The bulk configuration, represented as the EOS data, is shown in yellow circles, and lies in the relatively lower energy range. A noticeable aspect is that the fitted GAP model performs extremely well in the low-energy regime *i.e.* near equilibrium while showing equivalent performance in moderate and high-energy (non-equilibrium) regimes. On the other hand, the overall force errors are 128.42 meV/Å, which is quite low compared to state-of-the-art[23]. The bulk forces are uniformly distributed across the entire range of the force window (Fig. 3b). Under the low force window, the errors are reasonable, indicating that GAP can capture the low-strain, relatively stable local minima configurations. For the training set, the contribution from the size-dependent errors, as shown in the chord diagram in Fig. 3(c), most of the energy error in the fitting is caused by the small size clusters (less than 15 atoms), owing to the structural diversity and energy variance as shown in Fig. 1(b-c). The error contribution from the bulk structure is rather minimal, possibly because the training dataset contains only a few near-equilibrium bulk structures. Additionally, GAP tends to naturally capture bulk configurations as the correlation between the local structural environment and energetics becomes more consistent with increasing system size, approaching a bulk state. This observation is further supported by examining the error contributions of relatively large clusters (>30 atoms). For moderate-size clusters (<15 atoms and <30 atoms), the error contribution remains somewhat moderate compared to smaller and larger counterparts. Interestingly, the error contribution of force for different sizes and types of systems remains almost identical, possibly due to a tradeoff between distorted high-strain configurations and the presence of non-equilibrium local minima where energy variance is high, while less strain is present in the configurations. A similar trend can be observed in the test performance shown in Fig. 3 (d-e), where the overall energy error is 9.4 meV/atom, and the overall force error is 123.26 meV/Å. The contribution towards the overall test error from different cluster sizes and bulk datasets follows a similar trend as seen in the training set. In general, the contribution towards the overall error (energy or force) from bulk data is lower compared to cluster data (supplementary Fig. 3 (a-b)). Additionally, the lattice parameters and the overall magnitude and order of the cohesive energy of the bulk polymorphs reasonably agree with DFT values (Fig. 3 (c-d)).To independently assess the performance of our model and compare it against other models, we evaluated its performance on the validation data sampled using an EAM potential and the RL-MCTS method, as described in the cluster dataset section. Fig. 4(a, c) shows the energy and force predictions of our model. The GAP model performs exceptionally well in predicting both energy and forces, outperforming other machine learning potential models for Ni, such as GAP[52], SNAP[52,53], QSNAP[52], and the GAP[52], SNAP[52,53], QSNAP[52] model, as shown in Fig. 4(b, d). This is evident in both energy and force predictions.

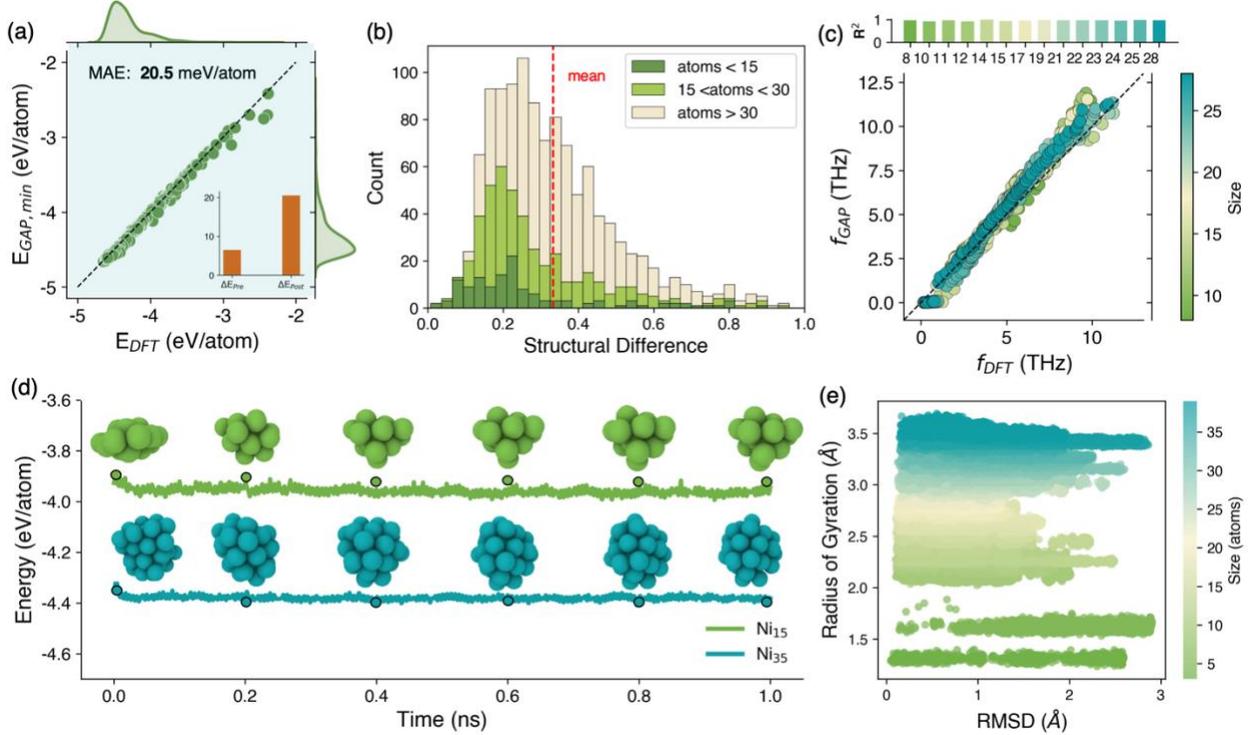

*Fig. 5. Stability and Dynamics of the GAP model on the Ni nanoclusters.* *(a) GAP minimized energy compared to DFT energies for size dependent DFT minima of Ni clusters as given in the QCD database. Also shows the MAE error in energies for pre- and post-relaxation (inset). (b) Structural similarity of the GAP minimized configuration from the QCD database with DFT minimized configurations. (c) Normal modes predicted by the GAP model compared with DFT-predicted normal modes for different sizes of clusters and corresponding $R^2$ of the predictions for different sizes. (d) Dynamic stability of the $Ni_{15}$ and $Ni_{35}$ nanoclusters at 300K. (e) Radius of Gyration vs Root Mean Square Deviation (RMSD) for the dynamic stability of clusters across different sizes.*

**Mapping the local minima and stability of nanoclusters:**

An important aspect of any potential model for low-dimensional materials is its ability to describe the various local minima in the energy landscape, encompassing both structural diversity and stability. This depends on the accuracy of the developed potentials as well as requires a comprehensive sampling of the configuration space to aid in the exploration of new minima. We begin by assessing the ability of our GAP model to capture the local minima in configurations of Ni, obtained from the Quantum Cluster Database (QCD)[44]. Initially, we calculate the unminimized Mean Absolute Error (MAE) of the energy and subsequently minimize the configurations using GAP. We then compare the minimized energies obtained with the GAP against those obtained through DFT (Fig. 5a). The minimized configuration exhibits an MAE error of 20.5 meV/atom, while that of the unminimized configurations is 6.5 meV/atom. This indicates our GAP model's capability to accurately capture the energetics of local minima configurations. Next, we compare the structural similarity of the GAP-minimized configurations with their DFT-minimized counterparts (Fig. 5b), utilizing a similarity measurement method described in the paper[54]. A general value of 0.3[54] indicates good structural similarity agreement. In our case, the mean values of the structural similarity scores closely approach 0.3. While smaller size clusters exhibit very good agreement with DFT configurations, the differences tend to increase with increments in the size of the configurations, as expected. Overall, all configurations align well with their DFT counterparts.

We further test the normal modes of the configurations across different sizes (with minimized energy error less than 3 meV/atom from DFT) which is considered a rigorous test of interatomic potentials and is typically done for single molecules and soft materials[23,55]. The normal modes predicted by the GAP are in very good agreement with the values predicted by DFT (Fig. 5c). Finally, we test the dynamic stability of the model on diverse topologies of configurations at room temperature. The dynamical stability of nanoclusters is among the most stringent tests that assess the structural integrity under certain conditions for prolonged periods of time[23]. We utilize LAMMPS[56] and perform canonical (NVT) ensemble simulations at 300 K for a total duration of 1ns. As seen in Fig. 5(d) for $Ni_{15}$ and $Ni_{35}$ clusters, the total cohesive energy as well as the structural integrity remain intact for the entire duration. Across different sizes, as shown in Fig. 5 (e), as the RMSD increases, which is expected due to diffusion within the clusters, the radius of gyration shows minimal change across different sizes. This suggests that there is little alteration in their shape, ensuring dynamic stability in the configurations.

**Transferability of GAP model from nanoscale to the bulk regime:**

One of our initial objectives was to make the GAP potential transferable to bulk systems, even though much of the training data comprised of properties of nanoclusters. To achieve this, we included bulk FCC configuration EOS data, enabling us to accurately capture lattice parameters, cohesive energy, and polymorph ordering (amongst FCC, BCC, HCP, and DC) (supplementary Fig. 3(c-d)). Additionally, we calculated the surface energies, EOS, and elastic modulus of the FCC configuration and compared them with DFT values (supplementary Fig. 4). Despite some degree of offset, our GAP model captures the surface energy ordering, though there is an under prediction (with a constant offset). The EOS closely matches its DFT counterpart (error ~4 meV/atom – supplementary Fig. 4b). The bulk and shear moduli predicted by the GAP model are 184 GPa and 60 GPa, respectively, compared to the DFT values of 198 GPa and 95 GPa[57]. This indicates a good agreement in the bulk modulus but an under prediction in the shear modulus. Although no explicit surface energy data were included during the fitting of the elastic and surface energy performance, the underperformance seems to stem from a trade-off between these properties and the model's accuracy for low-dimensional nanoclusters. A crucial and challenging property for any potential model is to capture the bulk thermodynamic properties such as the melting point. The experimental melting point of FCC Ni is ~1726 K[58]. As depicted in Fig. 6(a), the crystal-amorphous interface melting of FCC Ni indicates that the crystalline part completely amorphized at 1800 K, while the entire configuration entirely crystallizes at a temperature of 1750 K, suggesting that the melting point lies between 1750-1800 K. We simulated the interface at temperatures ranging from 1600 K to 1850 K and observed the potential energy variation of the configurations (supplementary Fig. 4(a)). These potential energy variations align with the observation in Fig. 6(a), suggesting that the melting point is around 1775 ± 25 K (including the 50 K fluctuation in the LAMMPS thermostat). This is very close to the experimental value, indicating that even with limited data on the energetics of bulk configurations (only EOS data) present in the training set, the GAP model can extrapolate well to capture most of bulk characteristics while retaining significant prediction accuracy in the cluster regime. Additionally, we compared the trends of GAP-predicted melting across different pressures with DFT values (extrapolated from the empirical model[59]) to assess the consistency of GAP's melting behavior with DFT (see Supplementary Fig. 4c). Our GAP model appears to align well with the trend, although the DFT methodology for calculating the melting temperature differs significantly from the approach adopted in our work. We further analyzed the size-dependent melting of Ni nanoparticles with our GAP model (Fig. 6b). using a Graph neural network model[33] (see methods and supplementary note 4). The GAP model demonstrates an increase in the melting point with increasing nanoparticle size. However, the predictions are higher than the experimental counterpart[14,60], but the GAP model successfully captures the trend in melting across sizes well beyond the training data it was originally fitted to. DFT is often known to underpredict surface energies compared to experimental values[61]. Our model shows a slight underprediction of surface energies relative to its DFT counterparts, leading to an overestimated cluster melting point compared to experimental results.

**Understanding the phase transition with graph neural networks**

Understanding phase transition mechanisms amongst various phases has been one of the prime goals of developing potential models for molecular simulations. The mobility of atoms driven by temperature gradients enables dynamic transformations of the material's surface and induces changes in the atomistic environments. Majority of previous efforts in analyzing molecular trajectories to gain mechanistic insights into phase behavior primarily relied on predefined mathematical formulation-based descriptors such as order parameters and the Lindemann index[14,19]. These descriptors work well in a relatively larger domain but tend to struggle as the dimensionality grows smaller, owing to mathematical boundaries. Recently, there have been attempts to leverage data instead of relying on predefined descriptors to explore the regime of nanoclusters[31,62]. Data holds the key to unraveling mechanistic insights into both the local environment and overall structural changes during phase stability and transitions. In this regard, Graph Neural Networks (GNNs) have been particularly successful in learning flexible representations by combining very fundamental low-level features (interatomic distances, bond angles, etc.) and yet producing a graph-based input that accurately maps to target-specific tasks. This caters to the limitations of static descriptors. One such workflow is the CEGANN model[33], which we have implemented to gain insights into the local melting behavior of $Ni_{420}$ nanoparticles (~2 nm diameter) (Fig. 6c-d). The specifics of training the CEGANN model are described in the method section. Fig. 6c shows frames of the nanoparticle trajectory at different temperatures and corresponding cross-section (at mid-plane) slices color-coded with the probability of being in the liquid phase as predicted by CEGANN. The bottom panel shows a histogram of all particle probabilities of being liquid across different temperatures. As early as 1000 K, we observe fluctuations in the in the local crystalline order with the probability varying from 0 (solid phase) to 1 (liquid phase), and these fluctuations disappear beyond 1250 K with almost all particles in the liquid phase, indicating early onset of local phase transitions in these nanoclusters. From the radial distribution (from the center of mass to the surface) of particle probabilities calculated for different frames in Fig. 6c, as shown in Fig. 6d, at around 650 K, only a few liquid-like particles reside close to the surface of the nanoparticles. As the temperature increases, the fluctuations indicative of melting transition starts to appear near the surface region (907 K), and these fluctuations subsequently propagate towards the center with further increments in temperature. As the temperature increases further (1278.8 K), most of the particles near the surface become liquid-like, except for only a few near the center, and there are fluctuations throughout the particle. At higher temperatures (1478.1 K), almost all of the particle's transition to the liquid phase. This validates surface pre-melting reported previously in Ni nanoclusters[21] and showcases the ability of the CEGANN model to capture intricacies of phase change learning solely from the data.

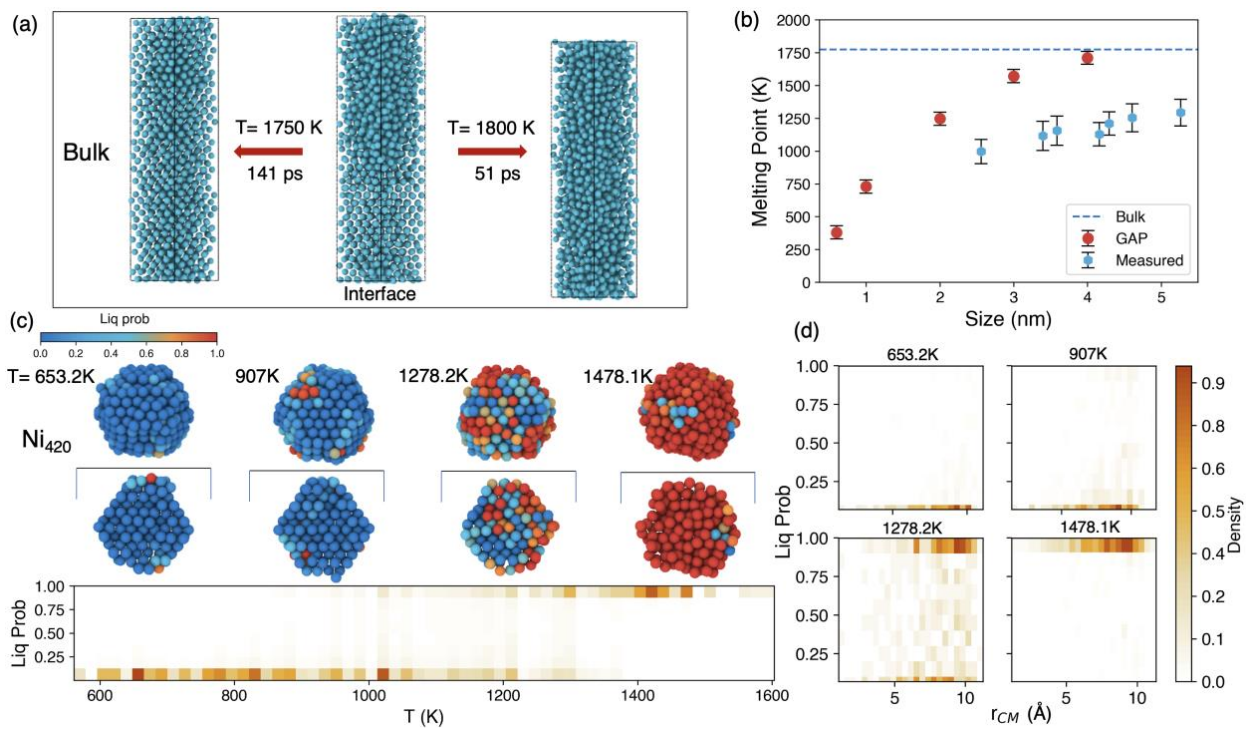

*Fig. 6. Melting characteristics of the GAP potential. (a) Interface melting of Bulk Ni structure. (b) Size-dependent melting of Ni Nanoparticles. (c) Phase change characteristics of $Ni_{420}$ (~2 nm diameter) nanoparticle. The upper section displays the nanoparticles and their cross-section (mid-plane) at different temperatures, with the particles color-coded according to their probability of being in the liquid phase as predicted by the CEGANN workflow. The bottom plot displays the liquid probability density of all the particles at different temperatures. (d) Shows the liquid probability density at different distances from the center of mass of the nanoparticle for different temperatures as depicted in (c).*

**Evaluating the thermos-mechanical stability of porous nanoparticles:**

Porosity in nanoparticles significantly increases their surface area and is useful in many practical applications. However, with the introduction of porosity, additional surface energy is introduced in the particle, altering its stability and temperature dynamics. Understanding the temperature-dependent stability of these nanoparticles is essential for gaining insight into the routes of their synthesis. The stability of these systems intricately depends on the inner and outer surface area. We implemented our GAP model to explore the stability of porous nanoparticles for different inner and outer diameters. The size ranges from 30-50 Å for the outer diameter and 20-30 Å for the inner diameter (Fig. 7b) with thickness varying from ~10-30 Å. Here, we define $T_c$ as the temperature at which the shell of the nanoparticle's collapses. We observe that for thicknesses <~15 Å, the nanoparticles collapse at a very (~ <300K) low temperature, while $T_c$ increases with thickness and reaches close to bulk melting eventually (Fig. 7b). Two main characteristics can be observed in the collapsed nanoparticles. For very small thicknesses (as depicted with 'A' in Fig. 7c), suffering an early collapse, and afterward, the nanoparticles recrystallize, which can be observed from the change in the radius of gyrations (Fig. 7b). For nanoparticles with relatively larger thickness, we only see two regimes with a stable shell and a collapsed shell (Fig. 7c, supplementary Fig. 5a).

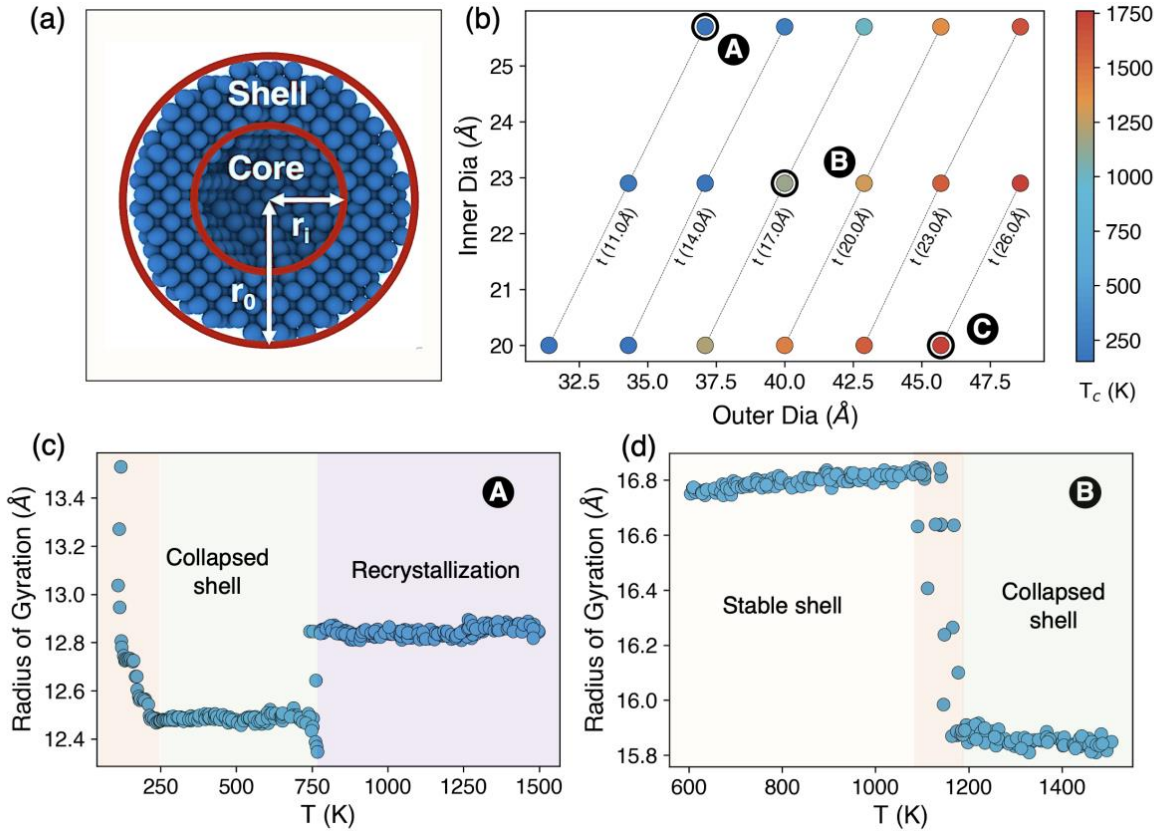

*Fig. 7. Thermo-mechanical stability of hollow nanoparticles of Ni. (a) Schematic of a porous nanoparticle. (b) Temperature of collapsing (Tc) for hollow nanoparticles of different inner and outer diameters and corresponding thickness(t). (c) Change in the radius of gyration of the nanoparticle as marked with 'A' in (a). (d) Change in the radius of gyration of the nanoparticle as marked with 'B' in (a).*

We further analyze the local environment of these two cases in Fig. 7 b-c using the trained CEGANN model, as depicted in Fig. 8. For the nanoparticle with smaller shell thickness (Fig. 8a-b), at lower temperatures (180 K) during the pre-collapse stage, most of the particles remain in the crystalline phase, with a few particles showing a liquid-like local environment at the outer surface. Interestingly, a collapse at a relatively high temperature (349 K) causes liquid-like particles at the surface to be trapped inside the core of the collapsed particle. This can also be observed in the distribution of the liquid-like particles at different temperatures (Fig. 8a) and in the distribution of the liquid probabilities of the particles radially (Fig. 8b). These trapped liquid-like particles cause the collapsed particle to crystallize from the inside as the temperature increases (693.2 K). Further increase in temperature causes liquid-like particles to appear on the surface, like a solid nanoparticle. For the porous nanoparticle with a relatively higher (~ >15 Å) thickness as depicted with 'B' in Fig. 7a, we see that the shell initially remains intact with increasing temperature. At a relatively high temperature (1190.5 K) (Fig. 8c), there is a collapse of the nanoparticle, forming a smaller hollow core. We also observe some liquid-like particles are present at the inner surface of the particles (Fig. 8d); however, unlike the previous case, there is no recrystallization observed as the temperature increases further. Instead, more liquid-like particles appear at the surface of the nanoparticle, indicating that it is approaching melting.

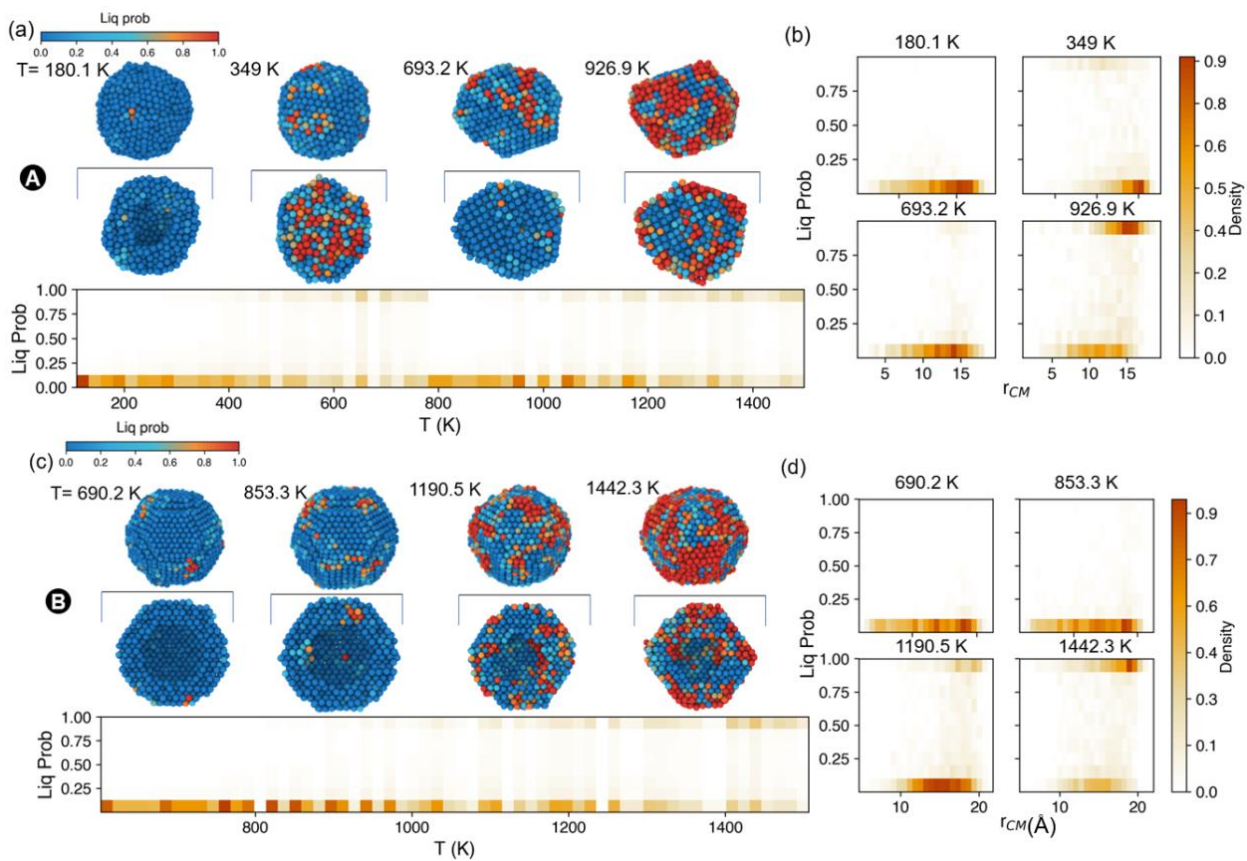

*Fig. 8. Evolution of hollow nanoparticles upon heating. (a) Evolution of the nanoparticles as in Fig. 7(a) - 'A' with increasing temperature. The upper section displays the nanoparticles and their cross-section at different temperatures, with the particles color-coded according to their probability of being in the liquid phase as predicted by the CEGANN workflow. The bottom plot displays the liquid probability density of all the particles at different temperatures. (b) Shows the liquid probability density at different distances from the center of mass of the hollow nanoparticle for different temperatures as depicted in (a). (c) Evolution of the nanoparticles as in Fig. 7(a) - 'B' with increasing temperature. (d) Shows the liquid probability density at different distances from the center of mass of the hollow nanoparticle for different temperatures as depicted in (c).*

**Discussion:**

ML potential models like GAP have recently shown remarkable accuracy and increased flexibility in mapping material properties across various systems, rivaling those of *ab initio* methods but at a fraction of the computational expense. Among the most intriguing material systems are zero-dimensional nanoclusters, distinguished by their unique surface properties and distinctive morphological traits, which render them applicable in numerous domains. In this study, we've employed a systematic approach to develop an ML potential model, GAP, tailored for a representative Ni nanocluster system. Starting with an understanding of the spatial and morphological impacts on system stability and energetics using a diverse training dataset, we developed a GAP model efficient in capturing the system's energetic and structural diversity. Additionally, we've explored the model's efficacy and error origins with varying system dimensionalities.

The advantage of GAP potential models lies in the flexibility of their functional form, allowing them to transform the problem of fitting the total energy of large systems into sum of smaller tasks. This flexibility enables accurate interpolation of the potential energy of structures in a broad training database, covering a wide variety of structures and phases. However, this flexibility often comes at the cost of the ability to interpolate to regions that are vastly different from what was encountered during the training of the model. The typical models, which are only functions of the local environment, make them short-range in nature. However, increasing the cutoff to some extent allows for long-range interactions to be accounted for. But this has its own pitfalls for these models, as increasing the cutoff for the construction of atomic environment vectors (such as SOAP in our case) results in a larger computational cost to capture physical interactions that slowly decay with separation distance, and the chemical configuration space grows with larger cutoff radii. We have attempted to assess the impact of these effects through a rigorous quality assessment of our developed model.

The model's quality assessment encompasses its capacity to represent local minima, vibrational modes, and system dynamics. A crucial aspect of this study involves testing the models' extrapolation capabilities. Consequently, with a small incorporation of representative bulk data, our model accurately predicted bulk properties, including lattice parameters, polymorph ordering, EOS, and elastic properties. This showcases the operational efficacies of these ML potential models across different regimes. It also emphasizes the fact that to achieve such a feat, the dataset should be well representative of the operational regime, and to ensure exploitability, the right representative candidates are needed at specific locations in representation spaces. Another aspect of this study emphasizes showcasing the efficacy of data-driven methods in complex phase change and stability problems and reveals insights that are difficult to obtain from traditional descriptors. To that end, we implemented a graph neural-based model to learn key characteristics that were crucial for understanding the melting and stability of porous nanoparticles. In general, the work emphasizes on approaches for developing ML potentials for nanoclusters that are model-agnostic and can be well generalized for the development and application of these potentials across various low-dimensional systems.

## Methods:

### The GAP workflow:

For parameterizing the GAP model, we employed the MAML package[42], which is built on top of the QUIP package[63]. All the cluster energies are scaled by subtracting the monomer energy(~0.1eV). The initial phase of training involved tuning hyperparameters, including the cutoff for constructing local neighborhoods, the sparsity parameters, and parameters related to constructing SOAP feature vectors, leading to multiple sets. Additionally, we opted to experiment with the inclusion of bulk Equation of State (EOS) data for FCC Ni in the training set to improve the extrapolative nature of the potential. The degree of emphasis placed on the energetics of bulk configurations during training was also considered a hyperparameter (supplementary Fig.2). The weight assigned to the cluster and bulk during fitting is simply proportional to the number of data points in each category. We keep the number of clusters fixed, and to adjust the cluster-to-bulk ratio, we add multiple copies of 63 unique bulk configurations. The primary objective was to keep the performance on the cluster dataset precise while capturing some of the bulk attributes. Subsequently, during the validation stage, performance metrics such as energy error, force error, error in lattice parameters, and the equation of states were evaluated and compared across different sets (refer to supplementary note 3). This process was iterated with additional sets of hyperparameters until convergence was achieved.

### Training of CEGANN model:

To train the CEGANN model, we generated a dataset of clusters ranging from diameters of 1-2nm, either entirely in the liquid phase or in the crystalline phase, across a diverse temperature range. Since this is a local classification task, the final number of particles in the dataset belonging to the liquid and crystalline phases was 38,636 and 52,640, respectively. We used an 80:20 split for training-validation data. The hyperparameters used to train the CEGANN model were the same as those used for the local classification of Zeolites as described in the paper[33], except for using 6 nearest neighbors, which seemed to yield good performance for our application. The validation accuracy reached ~1 after 50 epochs of training cycles (supplementary Fig. 6b), demonstrating the model's ability to map phase characteristics accurately. The confusion matrix (supplementary Fig. 6c) also shows the accuracy of the CEGANN model in segregating the two phases. It is worth noting that the CEGANN model was trained on both extremes of the phases, either crystalline or liquid, and extrapolates to regions in between, providing probabilities for individual particles being in either the liquid or crystalline phase.

**Molecular simulations specifics:**

The molecular simulations were carried out using LAMMPS, utilizing the QUIP package. All the cluster configurations were created using the Pymatgen[64] package, with a vacuum of 20 Å on all sides of the cluster configuration to ensure there is no periodic interaction. The dynamic stability simulations were computed using an NVT ensemble at 300K for a duration of 1 ns. While the two-phase melting simulations were conducted using an NPT ensemble at respective temperatures for a duration of 1 ns. The melting of nanoparticles and the stability simulations for the hollow nanoparticles were conducted using an NVT ensemble, with a heating rate of 0.2 K/ps.

**Data Availability:**

The training dataset used for this study is available from the authors upon reasonable request. The clusters from the quantum clusters database are available via https://muellergroup.jhu.edu/qcd/.

**Code availability:**

The developed GAP model is available at https://doi.org/10.5281/zenodo.11215543.

**Acknowledgement:**

This is based upon work supported by the US Department of Energy, Office of Science, Office of Basic Energy Sciences Data, Artificial Intelligence, and Machine Learning at DOE Scientific User Facilities program under Award Number 34532 (Digital Twins). This work was performed in part at the Center for Nanoscale Materials, which is a US Department of Energy of Science User facilities supported by the U.S. Department of Energy, Office of Science, Office of Basic Energy Sciences, under Contract No. DE-AC02-06CH11357. This work utilized National Energy Research Scientific Computing Center; a DOE Office of Science User Facility supported by the Office of Science of the U.S. Department of Energy under Contract No. DE-AC02-05CH11231. We also acknowledge the LCRC computing facilities at Argonne.

**Competing Interests:**

The authors declare no competing interests.

**Author contributions:**

S.K.R.S.S. conceived the project. S.B. developed the GAP models, conducted all calculations, and validated the model. S.B. performed all the analysis and wrote the manuscript with guidance from S.K.R.S.S. P.S.D. performed dynamic stability calculations. S.B. sampled the validation dataset, performed DFT calculations, and carried out the analysis. S.M. conducted the DFT computations for the generation of the training dataset and contributed toward its analysis, and provided valuable feedback on the manuscript. S.K.R.S.S. supervised and directed the overall project.

**Reference:**

(1) Banik, S.; Loefller, T.; Manna, S.; Chan, H.; Srinivasan, S.; Darancet, P.; Hexemer, A.; Sankaranarayanan, S. K. R. S. A Continuous Action Space Tree search for INverse desiGn (CASTING) framework for materials discovery. *npj Computational Materials* **2023,** *9* (1), 177.
(2) Wilcoxon, J. P.; Abrams, B. L. Synthesis, structure and properties of metal nanoclusters. *Chemical Society Reviews* **2006,** *35* (11), 1162.
(3) Chakraborty, I.; Pradeep, T. Atomically precise clusters of noble metals: emerging link between atoms and nanoparticles. *Chemical reviews* **2017,** *117* (12), 8208.
(4) Tang, Q.; Hu, G.; Fung, V.; Jiang, D.-e. Insights into interfaces, stability, electronic properties, and catalytic activities of atomically precise metal nanoclusters from first principles. *Accounts of chemical research* **2018,** *51* (11), 2793.
(5) Doye, J. P. K.; Wales, D. J. Global minima for transition metal clusters described by Sutton–Chen potentials. *New journal of chemistry* **1998,** *22* (7), 733.
(6) Doye, J. P. K.; Wales, D. J.; Berry, R. S. The effect of the range of the potential on the structures of clusters. *The Journal of chemical physics* **1995,** *103* (10), 4234.
(7) Cuenya, B. R. Synthesis and catalytic properties of metal nanoparticles: Size, shape, support, composition, and oxidation state effects. *Thin Solid Films* **2010,** *518* (12), 3127.
(8) Sau, T. K.; Rogach, A. L. Nonspherical noble metal nanoparticles: colloid-chemical synthesis and morphology control. *Advanced Materials* **2010,** *22* (16), 1781.
(9) Jaji, N.-D.; Lee, H. L.; Hussin, M. H.; Akil, H. M.; Zakaria, M. R.; Othman, M. B. H. Advanced nickel nanoparticles technology: From synthesis to applications. *Nanotechnology reviews* **2020,** *9* (1), 1456.
(10) Abdel Fattah, A. R.; Majdi, T.; Abdalla, A. M.; Ghosh, S.; Puri, I. K. Nickel nanoparticles entangled in carbon nanotubes: novel ink for nanotube printing. *ACS applied materials & interfaces* **2016,** *8* (3), 1589.
(11) Jiao, M.; Yao, Y.; Pastel, G.; Li, T.; Liang, Z.; Xie, H.; Kong, W.; Liu, B.; Song, J.; Hu, L. Fly-through synthesis of nanoparticles on textile and paper substrates. *Nanoscale* **2019,** *11* (13), 6174.
(12) Bibi, I.; Kamal, S.; Ahmed, A.; Iqbal, M.; Nouren, S.; Jilani, K.; Nazar, N.; Amir, M.; Abbas, A.; Ata, S. Nickel nanoparticle synthesis using Camellia Sinensis as reducing and capping agent:

(13) Nguyen, T. D.; Nguyen, C. C.; Tran, V. H. Molecular dynamics study of microscopic structures, phase transitions and dynamic crystallization in Ni nanoparticles. *RSC advances* **2017,** *7* (41), 25406.
(14) Rahbar, H.; Goudeli, E.; Kholghy, M. R. Sintering rate of nickel nanoparticles by molecular dynamics. *The Journal of Physical Chemistry C* **2023,** *127* (14), 6802.
(15) Jiang, S.; Zhang, Y.; Gan, Y.; Chen, Z.; Peng, H. Molecular dynamics study of neck growth in laser sintering of hollow silver nanoparticles with different heating rates. *Journal of Physics D: Applied Physics* **2013,** *46* (33), 335302.
(16) Schwartzberg, A. M.; Olson, T. Y.; Talley, C. E.; Zhang, J. Z. Synthesis, characterization, and tunable optical properties of hollow gold nanospheres. *The Journal of Physical Chemistry B* **2006,** *110* (40), 19935.
(17) Valencia, F. J.; Ramírez, M.; Varas, A.; Rogan, J. Understanding the stability of hollow nanoparticles with polycrystalline shells. *The Journal of Physical Chemistry C* **2020,** *124* (18), 10143.
(18) Yu, L.; Hu, H.; Wu, H. B.; Lou, X. W. Complex hollow nanostructures: synthesis and energy-related applications. *Advanced Materials* **2017,** *29* (15), 1604563.
(19) Neyts, E. C.; Bogaerts, A. Numerical study of the size-dependent melting mechanisms of nickel nanoclusters. *The Journal of Physical Chemistry C* **2009,** *113* (7), 2771.
(20) Li, B.; Li, J.; Su, X.; Cui, Y. Molecular dynamics study on structural and atomic evolution between Au and Ni nanoparticles through coalescence. *Scientific Reports* **2021,** *11* (1), 15432.
(21) Qi, Y.; Çağin, T.; Johnson, W. L.; Goddard Iii, W. A. Melting and crystallization in Ni nanoclusters: The mesoscale regime. *The journal of chemical physics* **2001,** *115* (1), 385.
(22) Pereyaslavets, L.; Kamath, G.; Butin, O.; Illarionov, A.; Olevanov, M.; Kurnikov, I.; Sakipov, S.; Leontyev, I.; Voronina, E.; Gannon, T. Accurate determination of solvation free energies of neutral organic compounds from first principles. *Nature communications* **2022,** *13* (1), 414.
(23) Manna, S.; Loeffler, T. D.; Batra, R.; Banik, S.; Chan, H.; Varughese, B.; Sasikumar, K.; Sternberg, M.; Peterka, T.; Cherukara, M. J. Learning in continuous action space for developing high dimensional potential energy models. *Nature communications* **2022,** *13* (1), 368.
(24) Jinnouchi, R.; Hirata, H.; Asahi, R. Extrapolating energetics on clusters and single-crystal surfaces to nanoparticles by machine-learning scheme. *The Journal of Physical Chemistry C* **2017,** *121* (47), 26397.
(25) Zeni, C.; Rossi, K.; Glielmo, A.; Baletto, F. On machine learning force fields for metallic nanoparticles. *Advances in Physics: X* **2019,** *4* (1), 1654919.
(26) Zhang, L.; Csányi, G.; Van Der Giessen, E.; Maresca, F. Atomistic fracture in bcc iron revealed by active learning of Gaussian approximation potential. *npj Computational Materials* **2023,** *9* (1), 217.
(27) Maresca, F.; Dragoni, D.; Csányi, G.; Marzari, N.; Curtin, W. A. Screw dislocation structure and mobility in body centered cubic Fe predicted by a Gaussian Approximation Potential. *npj Computational Materials* **2018,** *4* (1), 69.
(28) Unruh, D.; Meidanshahi, R. V.; Goodnick, S. M.; Csányi, G.; Zimányi, G. T. Gaussian approximation potential for amorphous Si: H. *Physical Review Materials* **2022,** *6* (6), 065603.
(29) Bartók, A. P.; Payne, M. C.; Kondor, R.; Csányi, G. Gaussian approximation potentials: The accuracy of quantum mechanics, without the electrons. *Physical review letters* **2010,** *104* (13), 136403.
(30) Szlachta, W. J.; Bartók, A. P.; Csányi, G. Accuracy and transferability of Gaussian approximation potential models for tungsten. *Physical Review B* **2014,** *90* (10), 104108.
(31) Zeni, C.; Rossi, K.; Pavloudis, T.; Kioseoglou, J.; de Gironcoli, S.; Palmer, R. E.; Baletto, F. Data-driven simulation and characterisation of gold nanoparticle melting. *Nature Communications* **2021,** *12* (1), 6056.

(32) Li, Z. H.; Truhlar, D. G. Nanosolids, slushes, and nanoliquids: characterization of nanophases in metal clusters and nanoparticles. *Journal of the American Chemical Society* **2008**, *130* (38), 12698.
(33) Banik, S.; Dhabal, D.; Chan, H.; Manna, S.; Cherukara, M.; Molinero, V.; Sankaranarayanan, S. K. R. S. CEGANN: Crystal Edge Graph Attention Neural Network for multiscale classification of materials environment. *npj Computational Materials* **2023**, *9* (1), 23.
(34) Banik, S.; Balasubramanian, K.; Manna, S.; Derrible, S.; Sankaranarayananan, S. K. R. S. Evaluating generalized feature importance via performance assessment of machine learning models for predicting elastic properties of materials. *Computational Materials Science* **2024**, *236*, 112847.
(35) Zeni, C.; Rossi, K.; Glielmo, A.; Fekete, Á.; Gaston, N.; Baletto, F.; De Vita, A. Building machine learning force fields for nanoclusters. *The Journal of chemical physics* **2018**, *148* (24).
(36) Wang, N.; Huang, S. Molecular dynamics study on magnesium hydride nanoclusters with machine-learning interatomic potential. *Physical Review B* **2020**, *102* (9), 094111.
(37) Pihlajamaki, A.; Hamalainen, J.; Linja, J.; Nieminen, P.; Malola, S.; Karkkainen, T.; Hakkinen, H. Monte Carlo simulations of $Au_{38}(SCH_3)_{24}$ nanocluster using distance-based machine learning methods. *The Journal of Physical Chemistry A* **2020**, *124* (23), 4827.
(38) Xiao, L.; Tollberg, B.; Hu, X.; Wang, L. Structural study of gold clusters. *The Journal of chemical physics* **2006**, *124* (11).
(39) Zhao, J.; Shi, R.; Sai, L.; Huang, X.; Su, Y. Comprehensive genetic algorithm for ab initio global optimisation of clusters. *Molecular Simulation* **2016**, *42* (10), 809.
(40) Klawohn, S.; Darby, J. P.; Kermode, J. R.; Csányi, G.; Caro, M. A.; Bartók, A. P. Gaussian approximation potentials: Theory, software implementation and application examples. *The Journal of Chemical Physics* **2023**, *159* (17).
(41) Bartók, A. P.; Kondor, R.; Csányi, G. On representing chemical environments. *Physical Review B* **2013**, *87* (18), 184115.
(42) Chen, C.; Ong, S. P. Maml - materials machine learning package. *GitHub repository* **2020**.
(43) Loeffler, T. D.; Patra, T. K.; Chan, H.; Cherukara, M.; Sankaranarayanan, S. K. R. S. Active learning the potential energy landscape for water clusters from sparse training data. *The Journal of Physical Chemistry C* **2020**, *124* (8), 4907.
(44) Manna, S.; Wang, Y.; Hernandez, A.; Lile, P.; Liu, S.; Mueller, T. A database of low-energy atomically precise nanoclusters. *Scientific Data* **2023**, *10* (1), 308.
(45) Wales, D. J.; Doye, J. P. K.; Dullweber, A.; Hodges, M. P.; Naumkin, F. Y.; Calvo, F.; Hernández-Rojas, J.; Middleton, T. F., 2001.
(46) Langenberg, A.; Hirsch, K.; Ławicki, A.; Zamudio-Bayer, V.; Niemeyer, M.; Chmiela, P.; Langbehn, B.; Terasaki, A.; Issendorff, B. v.; Lau, J. T. Spin and orbital magnetic moments of size-selected iron, cobalt, and nickel clusters. *Physical Review B* **2014**, *90* (18), 184420.
(47) Chaves, A. S.; Piotrowski, M. J.; Da Silva, J. L. F. Evolution of the structural, energetic, and electronic properties of the 3d, 4d, and 5d transition-metal clusters (30 $TM_n$ systems for n= 2–15): a density functional theory investigation. *Physical chemistry chemical physics* **2017**, *19* (23), 15484.
(48) Zhang, Y.; Ashcraft, R.; Mendelev, M. I.; Wang, C. Z.; Kelton, K. F. Experimental and molecular dynamics simulation study of structure of liquid and amorphous $Ni_{62}Nb_{38}$ alloy. *The Journal of chemical physics* **2016**, *145* (20).
(49) Zeng, C.; Chen, Y.; Iida, K.; Nobusada, K.; Kirschbaum, K.; Lambright, K. J.; Jin, R. Gold quantum boxes: on the periodicities and the quantum confinement in the $Au_{28}$, $Au_{36}$, $Au_{44}$, and $Au_{52}$ magic series. *Journal of the American Chemical Society* **2016**, *138* (12), 3950.
(50) Liu, G. F.; Sroubek, Z.; Yarmoff, J. A. Detection of quantum confined states in Au nanoclusters by alkali ion scattering. *Physical review letters* **2004**, *92* (21), 216801.
(51) Anstine, D. M.; Isayev, O. Machine learning interatomic potentials and long-range physics. *The Journal of Physical Chemistry A* **2023**, *127* (11), 2417.

(52) Zuo, Y.; Chen, C.; Li, X.; Deng, Z.; Chen, Y.; Behler, J. r.; Csányi, G.; Shapeev, A. V.; Thompson, A. P.; Wood, M. A. Performance and cost assessment of machine learning interatomic potentials. *The Journal of Physical Chemistry A* **2020,** *124* (4), 731.

(53) Li, X.-G.; Hu, C.; Chen, C.; Deng, Z.; Luo, J.; Ong, S. P. Quantum-accurate spectral neighbor analysis potential models for Ni-Mo binary alloys and fcc metals. *Physical Review B* **2018,** *98* (9), 094104.

(54) Li, X.-T.; Yang, X.-B.; Zhao, Y.-J. Geometrical eigen-subspace framework based molecular conformation representation for efficient structure recognition and comparison. *The Journal of Chemical Physics* **2017,** *146* (15).

(55) Bahar, I.; Lezon, T. R.; Bakan, A.; Shrivastava, I. H. Normal mode analysis of biomolecular structures: functional mechanisms of membrane proteins. *Chemical reviews* **2010,** *110* (3), 1463.

(56) Thompson, A. P.; Aktulga, H. M.; Berger, R.; Bolintineanu, D. S.; Brown, W. M.; Crozier, P. S.; In't Veld, P. J.; Kohlmeyer, A.; Moore, S. G.; Nguyen, T. D. LAMMPS-a flexible simulation tool for particle-based materials modeling at the atomic, meso, and continuum scales. *Computer Physics Communications* **2022,** *271*, 108171.

(57) Jain, A.; Ong, S. P.; Hautier, G.; Chen, W.; Richards, W. D.; Dacek, S.; Cholia, S.; Gunter, D.; Skinner, D.; Ceder, G. Commentary: The Materials Project: A materials genome approach to accelerating materials innovation. *APL materials* **2013,** *1* (1).

(58) Jordan, L.; Swanger, W. H. The properties of pure nickel. *Bur. Stand. J. Res* **1930,** *5* (6), 1291.

(59) Pozzo, M.; Alfe, D. Melting curve of face-centered-cubic nickel from first-principles calculations. *Physical Review B—Condensed Matter and Materials Physics* **2013,** *88* (2), 024111.

(60) van Teijlingen, A.; Davis, S. A.; Hall, S. R. Size-dependent melting point depression of nickel nanoparticles. *Nanoscale Advances* **2020,** *2* (6), 2347.

(61) Holec, D.; Dumitraschkewitz, P.; Vollath, D.; Fischer, F. D. Surface energy of Au nanoparticles depending on their size and shape. *Nanomaterials* **2020,** *10* (3), 484.

(62) Rapetti, D.; Delle Piane, M.; Cioni, M.; Polino, D.; Ferrando, R.; Pavan, G. M. Machine learning of atomic dynamics and statistical surface identities in gold nanoparticles. *Communications Chemistry* **2023,** *6* (1), 143.

(63) Csányi, G.; Winfield, S.; Kermode, J.; Payne, M. C.; Comisso, A.; De Vita, A.; Bernstein, N. Expressive programming for computational physics in Fortran 950+. *Newsletter of the Computational Physics Group* **2007**, 1.

(64) Ong, S. P.; Richards, W. D.; Jain, A.; Hautier, G.; Kocher, M.; Cholia, S.; Gunter, D.; Chevrier, V. L.; Persson, K. A.; Ceder, G. Python Materials Genomics (pymatgen): A robust, open-source python library for materials analysis. *Computational Materials Science* **2013,** *68*, 314.

# Supplementary information

[1] Center for Nanoscale Materials, Argonne National Laboratory, Lemont, Illinois 60439.

[2] Department of Mechanical and Industrial Engineering, University of Illinois, Chicago, Illinois 60607.

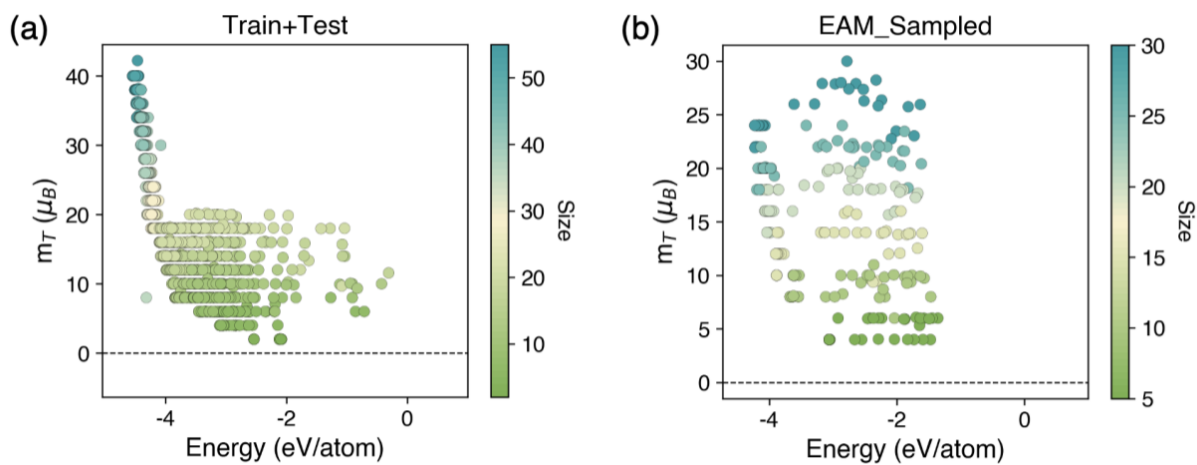

***Supplementary Fig.1*** *(a) Total magnetic moment of configurations used as training and testing data for different sizes. (b) The total magnetic moment of newly sampled configurations using an EAM model[1] model for further assessment of the performance of the developed GAP model.*

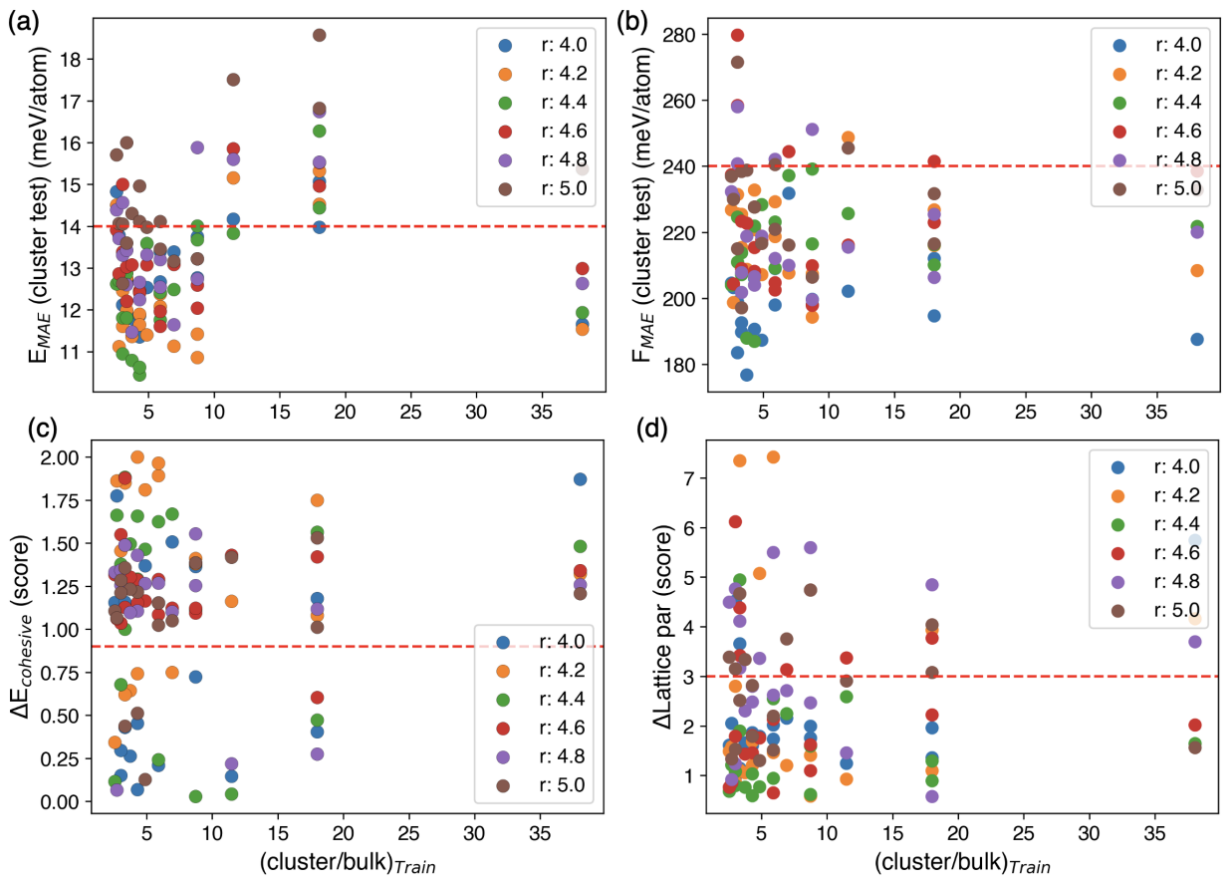

*Supplementary Fig.2. Determining the ratio weight of bulk data compared to cluster data in the training set:* (a) The ratio weight of cluster data to bulk data in the training set and the subsequent Mean Absolute Error (MAE) in energies on the clusters of the test set for different GAP models obtained. (b) The ratio weight of cluster data to bulk data in the training set and the subsequent MAE error in forces on the clusters of the test set for different GAP models obtained. (c) A score giving accuracy in the predictability of cohesive energies of ground state polymorphs, i.e., FCC, BCC, HCP, and DC (Diamond Cubic), for different ratios of cluster to bulk weight in the training set. (d) A score giving the accuracy in the predictability of lattice parameters of ground state polymorphs, i.e., FCC, BCC, HCP, and DC (Diamond Cubic), for different ratios of cluster to bulk weight in the training dataset. The red line indicates the selection cutoff for both cases, based on which common GAP models are selected for further evaluation. The red lines indicate the cutoff used for each of the above criteria for the selection of the model.

## Supplementary Note 1:

**DFT methodology for the generation of the datasets:**

The Vienna Ab-initio Software Package (VASP)[2] was utilized, employing the Perdew-Burke-Eznerhof (PBE) exchange-correlation functional[3], for all density functional theory (DFT) calculations to evaluate energy and forces of clusters in both the training and test datasets. These DFT calculations included spin polarization. The projector augmented wave (PAW) potentials utilized in these calculations. Each calculation employed a single k-point located at the center of the Brillouin zone. Gaussian smearing with a width of 0.001 eV was applied to set partial occupancies. The convergence criteria for the electronic self-consistent iteration and the ionic relaxation loop were set at 0.1 meV and 1 meV per cluster, respectively. A dense k-point grid was utilized, defined by the product of the number of atoms in the primitive cell and the number of k-points, aiming for approximately 1000 points. The DFT calculations were employed to determine cohesive energies and lattice parameters. All of our VASP calculations for energy and forces employed spin-polarized methods, with various initial MAGMOM values tested, which eventually converged to a ferromagnetic state, similar to Ref. 4[4] (Supplementary Fig.1).A relatively high tolerance of $10e^{-6}$ eV for energy convergence was employed. Elastic constants were determined using suitable lattice distortions implemented in the VASP finite difference methods.

**Supplementary Note 2:**

The Metropolis simulation was conducted over 5,000 Monte Carlo (MC) cycles at 300K, starting with the initial structure randomly selected from a dimer. Subsequently, Nested Ensemble simulations were carried out for an additional 500,000 cycles, with configurations saved after each 1,000 cycles. In both simulation methods, a cycle consisted of N moves, where N was set to the number of atoms in the given simulation. This approach facilitated the generation of cluster configurations spanning a wide range of energy windows. The number of atoms for each element varied from 5 to 30. To ensure proper clustering during the MC simulations, a Stillinger Cluster Criteria[5,6] was enforced to maintain connectivity. Any configuration that risked breaking a cluster was rejected in favor of the previous position. For each system, the cluster criteria distance was set to the first neighbor distance between minimized atoms, with an additional tolerance of 0.5 Å to accommodate structural flexibility.

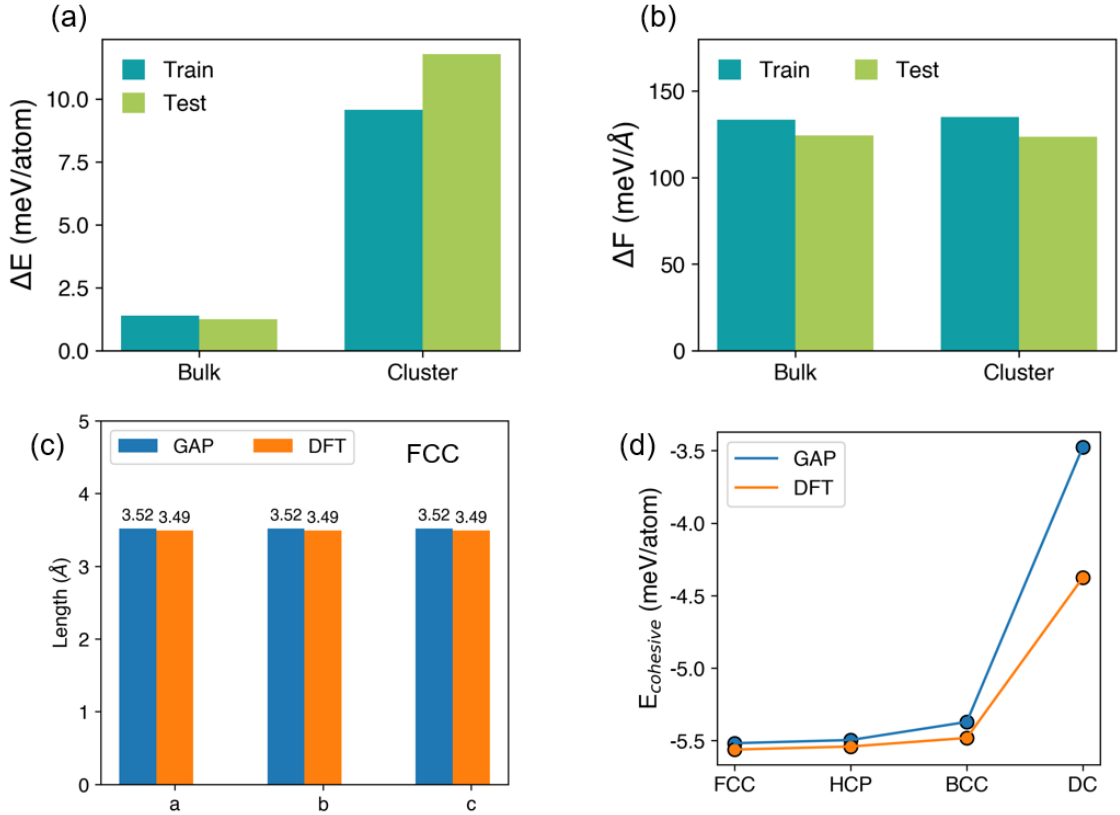

*Supplementary Fig.3: The error contribution from bulk and cluster configurations to the overall errors is depicted in Fig. 1 of the main manuscript: (a) Overall Energy error contribution from bulk and cluster configurations on the training and test datasets. (b) Overall Force error contribution from bulk and cluster configurations on the training and test datasets. (c) Lattice parameter of the FCC polymorph as predicted by the GAP compared to DFT. (d) Ordering and magnitude of the cohesive energy as predicted by the GAP compared to the respective DFT values for FCC, BCC, HCP, and DC bulk polymorphs of Ni.*

## Supplementary Note 3:

To test the fraction of bulk Equations of State (EOS) data to be included in the training set to accurately capture bulk attributes such as cohesive energy and the respective ordering of ground state polymorphs (FCC, BCC, HCP, and DC (Diamond Cubic)), as well as the accuracy in predicting their lattice parameters while keeping the prediction accuracy on the cluster dataset reasonable, we experimented with different ratio weights of cluster data to bulk data in the training set, as depicted in Supplementary Fig. 2. The score provides a quantitative measure of relative accuracy in predicting cohesive energy while maintaining the ordering across the polymorphs (Supplementary Fig. 2(c)).

$$\Delta E_{Cohesive} = \|E_{polymorhs, predicted} - E_{polymorps, target}\| + S_{order}$$

where $S_{order}$ is the ordering score, where 0 indicates ordered and 1 indicates misordering. Additionally, the score for predicting the accuracy of lattice parameters is given by (Supplementary Fig. 2(d)):

$$\Delta lattice = 4*\Delta\|L_{FCC}\| + 3*\Delta\|L_{BCC}\| + 2*\Delta\|L_{HCP}\| + \Delta\|L_{DC}\|$$

where $\|L\|$ is the norm of target and predicted lattice parameters. Maximum weight is given to the prediction of FCC lattice parameters. Finally, all the GAP models having errors on the test cluster dataset for energy (Supplementary Fig. 2(a)) less than 14 meV/atom, force (Supplementary Fig. 2(b)) less than 240 meV/Å, cohesive energy error less than 0.9, and lattice error score less than 3 are chosen for further evaluation.

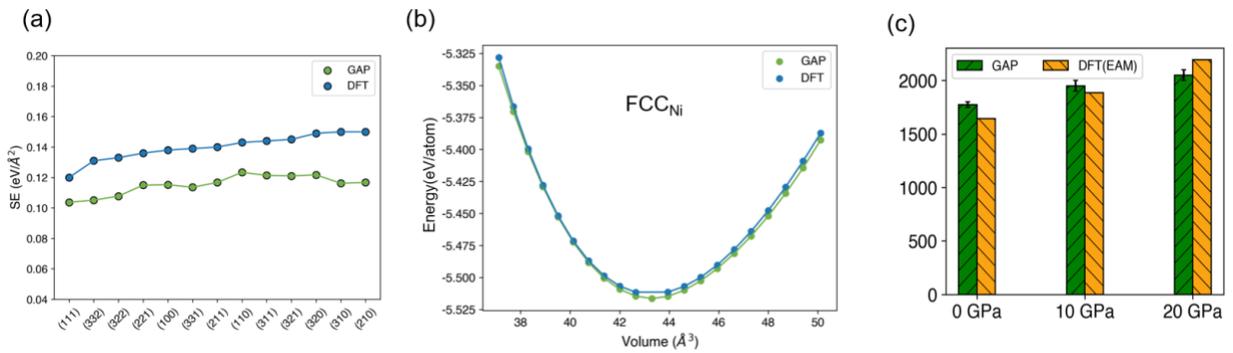

*Supplementary Fig.4: Performance of the GAP potential for bulk properties: (a) The surface energies predicted by the GAP potential compared with the DFT counterpart. (b) Shows the equations of state (EOS) of FCC configuration predicted by the GAP potential vs. the DFT-predicted equations of state. (c) Comparison of bulk melting point predicted by our GAP model with its DFT (extrapolated from EAM model[1]) counterpart at different pressures.*

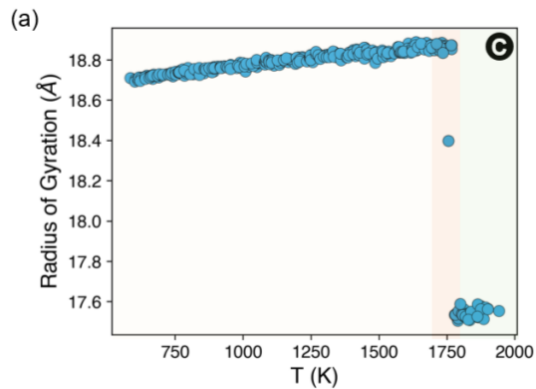

*Supplementary Fig. 5. Stability of hollow nanoparticles of Ni. (a) Change in the radius of gyration of the nanoparticle as marked with 'C' in main manuscript Fig 5 (a).*

## Supplementary Note 4:

To determine the melting point of a nanoparticle, we first assess the probability of each particle belonging to either the liquid or crystalline phase using the trained CEGANN- model as described in the main manuscript. Subsequently, we determine whether each particle is in the liquid phase by considering the maximum probability of particles that has more than 90% probability of being liquid. We then monitor the total number of liquid particles in the cluster at various temperatures (supplementary Fig. 6 d-f). The temperature at which the total number of liquid particles in the cluster surpasses 60% of the total particles is designated as the melting point of the nanoparticle.

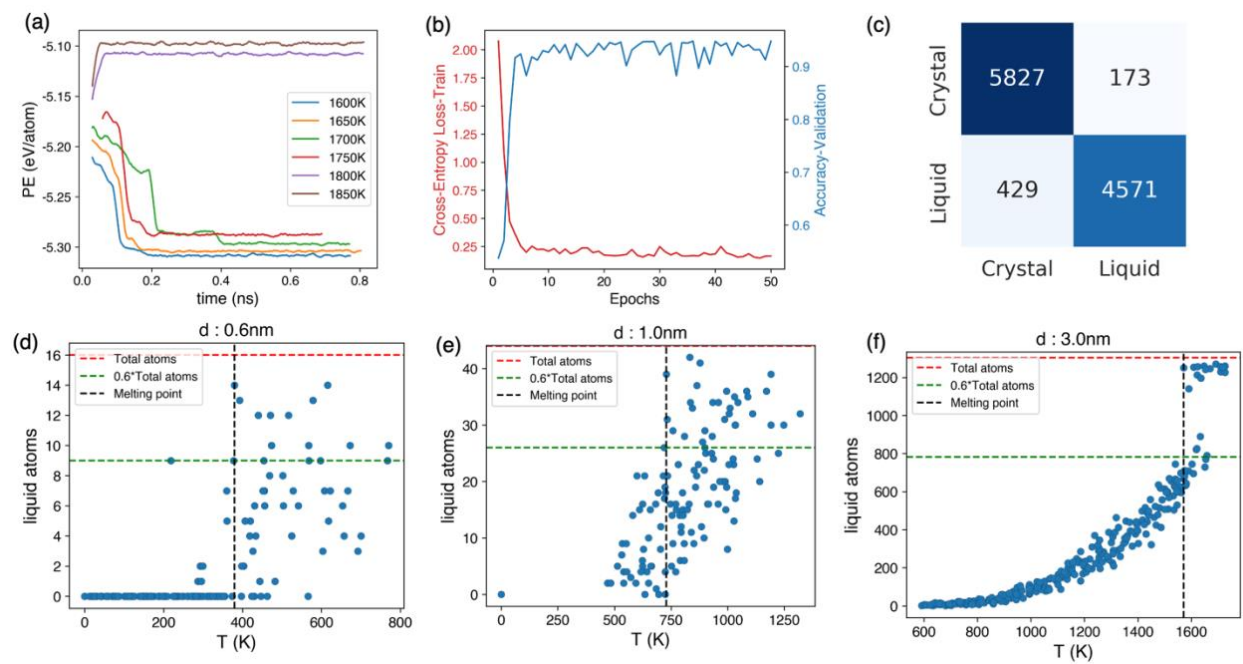

*Supplementary Fig.6: Melting and phase change with the GAP potential: (a) The potential energy of the interface-bulk configuration at different temperatures. (b) The train-test performance of the CEGANN workflow trained on the nanoclusters of liquid and crystalline phases. (c) The confusion matrix on the validation set in terms of the number of particles in each class as predicted by the trained CEGANN model in (b). (d-f) The evolution of the total number of particles belonging to the liquid phase with temperature for nanoparticle diameters of 0.6, 1, and 3 nm, and determination of the melting point based on the count of liquid nanoparticles once it has reached 60% of the total particles in the system.*

**Reference:**

1	Zhang, Y., Ashcraft, R., Mendelev, M. I., Wang, C. Z. & Kelton, K. F. Experimental and molecular dynamics simulation study of structure of liquid and amorphous Ni62Nb38 alloy. *The Journal of chemical physics* **145** (2016).
2	Kresse, G. & Hafner, J. Ab initio molecular dynamics for liquid metals. *Physical review B* **47**, 558 (1993).
3	Ernzerhof, M. & Scuseria, G. E. Assessment of the Perdew–Burke–Ernzerhof exchange-correlation functional. *The Journal of chemical physics* **110**, 5029-5036 (1999).
4	Manna, S. *et al.* A database of low-energy atomically precise nanoclusters. *Scientific Data* **10**, 308 (2023).
5	Manna, S. *et al.* Learning in continuous action space for developing high dimensional potential energy models. *Nature communications* **13**, 368 (2022).
6	Stillinger Jr, F. H. Rigorous basis of the frenkel-band theory of association equilibrium. *The Journal of Chemical Physics* **38**, 1486-1494 (1963).